\documentclass{aa}  

\usepackage{graphicx}
\usepackage{txfonts}
\usepackage[colorlinks=true,citecolor=blue,linkcolor=blue]{hyperref}
\usepackage{subfigure}
\usepackage{subcaption}

\bibpunct{(}{)}{;}{a}{}{,}
\defcitealias{andersson24starclusters}{A24}

\usepackage{orcidlink}

\usepackage{etoolbox}
\makeatletter
\AtBeginDocument{\@ifpackageloaded{lineno}{\nolinenumbers\renewcommand{\linenumbers}{} }{}}
\makeatother

\begin{document}

   \title{A potential observational test to disentangle the companion of binary millisecond pulsar PSR J1928+1815}

\titlerunning{A potential observational test to disentangle the companion PSR J1928+1815}
\authorrunning{Wen-Cong Chen}
   \author{Wen-Cong Chen\inst{1}\orcidlink{0000-0002-0785-5349}
                  }
   \institute{School of Science, Qingdao University of Technology, Qingdao 266525, People's Republic of China \email{chenwc@pku.edu.cn}}

   \date{Received ; accepted }
 
  \abstract  
   {Recently, the binary millisecond pulsar (MSP) PSR J1928+1815 (J1928) was reported to be experiencing regular radio eclipses. To produce the radio pulse emission and regular radio eclipses, this source was proposed to be most likely an MSP with a He star companion. However, the possibility that the companion is a white dwarf (WD) cannot be completely ruled out. Considering the uncertainty of interstellar dust extinction, it is difficult to confirm the nature of the companion star in the near-infrared and optical bands. Based on the measured mass function and orbital inclination angle inferred by the radio eclipses, the gravitational radiation can produce a period derivative of $-(6.8-9.6)\times10^{-13}~\rm s\,s^{-1}$ when the companion star of J1928 is an extremely massive WD ($1.4~M_\odot$). Therefore, its companion star can be identified as a He star with a mass higher than $1.4~M_\odot$ if the measured period derivative exceeds $-9.6\times10^{-13}~\rm s\,s^{-1}$. However, the expansion of a He star could cause a transfer of angular momentum from the orbital motion to the spin of the He star through a tidal interaction. The loss of orbital angular momentum results in a negative orbital period derivative, which produces an extra-contribution to the period derivative due to gravitational radiation. If the companion of J1928 is a $1.0-1.6~M_\odot$ He star, the predicted period derivative is in the range of $-(4.9-13.5)\times10^{-13}~\rm s\,s^{-1}$. In particular, the tidal interaction could help to confirm a He star with a mass of $1.35-1.5~M_\odot$. We also constrain the maximum evaporation efficiency of spin-down luminosity to be $0.006$ when the companion star of J1928 is a He star.}


   \maketitle
\section{Introduction}\label{sec:intro}

Millisecond pulsars (MSPs) are a subpopulation of radio pulsars with a spin period shorter than 30 ms and a weak magnetic field ($\sim10^8-10^9~\rm G$). MSPs are generally thought to be spun up to a millisecond period by accretion of material and angular momentum from a donor star in low-mass X-ray binaries \citep[LMXBs,][]{alpa82,bhat91}. The discovery of accreting MSPs \citep{wijn98} and transitional MSPs including PSR J1023+0038 \citep{arch09,patr14,stap14}, IGR J18245-2452 \citep{papi13}, and XSS J12270-4859 \citep{mart10,mart13} provided strong support for the aforementioned recycling theory.

In binary MSPs, there exist two classes of spider MSPs with regular radio eclipses and very compact orbits \citep[orbital periods are shorter than 1 day,][]{robe13}. According to companion masses, spider MSPs divide into black widows with brown dwarf companions \citep[with a mass of $0.01-0.05~M_\odot$,][]{fruc88,stap96} or redbacks with main-sequence (MS) companions \citep[with a mass of $0.1-0.8~M_\odot$,][]{robe13}. A detailed binary evolution model performed by \cite{chen13} indicated that redbacks cannot evolve into black widows; thus, they are two distinct populations.

During the Galactic Plane Pulsar Snapshot (GPPS) survey of Five-hundred-meter Aperture Spherical radio Telescope (FAST), PSR J1928+1815 (J1928) was identified to have a spin period of 10.55 ms and be in a binary system with an orbital period of 3.60 hours \citep{han21}. By the 4.5 years of follow-up pulsar timing observations using FAST, this MSP was detected to experience three regular long radio eclipses during the orbital phase range of 0.18 to 0.35 \citep{yang25}. According to the observed mass function of $0.2342~M_\odot$, the minimum companion mass is about $1.0~M_\odot$. A counterpart cannot be found by searching optical and infrared catalogs, in which the severest constraint is the archive data of the infrared K-band at $2.2~\mu \rm m$ \citep{yang25}. A population synthesis simulation using the modified code BSE indicated that the Galaxy has dozens and one percent PSR J1928+1815-like systems when the companion is a He star or proto-white dwarf (WD), respectively \citep{yang25}. Based on the simulated result of population synthesis and no detection in the optical and infrared bands, \cite{yang25} proposed that J1928 is most likely to be a post-common envelope (CE) system, in which the companion star is a stripped helium (He) star with a mass of $1.0-1.6~M_\odot$.

The formation of the MSP J1928 remains controversial. \cite{yang25} proposed that the millisecond period originated from super-Eddington accretion in the CE phase, during which neutrinos carried away a large amount of released gravitational potential energy. A detailed model performed by \cite{guo25} indicated that J1928 may undergo a mass-transfer phase prior to the current detached stage. \cite{deng25} found that the short spin period of J1928 may arise from super-Eddington accretion through Roche lobe overflow prior to the CE phase. However, \cite{nie26} demonstrated that neutron stars (NSs) can be spun up to a millisecond period by accretion in a CE decoupling phase.

\cite{gong25} performed a near-infrared (NIR) imaging observation with limiting magnitudes of $J = 23.7$ and $H = 22.2$ for J1928, which are both 1.6 mag deeper than the predicted magnitudes in these two bands for a modeled naked He star with a mass of $1.0~M_\odot$ after extinction. However, they cannot detect any signal from this source. As an alternative model, \cite{gong25} proposed that J1928 is a peculiar black widow system with a heavy ablated WD. Recently,  \cite{naga26} did not discover any source consistent with the radio position of J1928 down to a $5\sigma$ limit of $K_{\rm s}\approx21.3$ through deep near-infrared imaging using Keck/NIRC2 with laser guide star adaptive optics. Therefore, they proposed that the companion star of J1928 is a young, massive WD, which can drive a strong wind to produce radio eclipses on a timescale of $10^4-10^5~\rm yr$. However, the  LAMOST-LRS survey found that a stable, long-lived circumstellar environment could persist around those post-CE systems \citep{li25}. The circumstellar environment could produce a wide range of extinction values up to four magnitudes in the $K_{\rm s}$ band \citep{lewi23}. Considering the high uncertainty of the circumstellar extinction, the possibility that the companion star of J1928 is a He star can still not be fully ruled out.

According to the regular radio eclipses, J1928 should belong to spider pulsars. The minimum mass of the companion rules out the possibility of a black widow. If the companion is a main-sequence star, it will fill the Roche lobe in the present compact orbit system and transfer mass onto the pulsar, quenching the radio pulse emission \citep{yang25}. Therefore, J1928 is a rare spider pulsar. Currently, the mysterious companion of J1928 has two possibilities: a He star or a WD. If the companion is a He star, J1928 is a unique binary MSP with a He star companion. Conversely, it is a spider pulsar with a WD companion, which has a probability much lower than that of a He star in stellar evolution. As a consequence, J1928 is a precious fossil in understanding stellar and binary evolutionary theory.

In this work, we propose a potential observational test to disentangle the companion nature of J1928. In Section 2, the mass function of J1928 is used to constrain the NS mass. We investigate the orbital evolution of J1928 under two companion cases, including a WD and a He star, and predict their orbital-period derivatives in Sections 3 and 4, respectively. In Section 5, we give a brief discussion and summary.

\section{Constraining the NS mass of J1928}

Employing pulsar timing observations using FAST, the mass function of J1928 was derived as follow \citep{yang25}
\begin{equation}
f(M_{\rm ns},M_{\rm c},i)=\frac{(M_{\rm c}\sin i)^3}{(M_{\rm ns}+M_{\rm c})^2}=0.2342~M_\odot,
\end{equation}
where $M_{\rm ns}$, $M_{\rm c}$, and $i$ are the NS mass, the companion mass, and the orbital inclination angle, respectively. Therefore, the NS mass of J1928 can be constrained as
\begin{equation}
M_{\rm ns}=\sqrt{\frac{(M_{\rm c}\sin i)^3}{0.2342~M_\odot}}-M_{\rm c}.
\end{equation}
J1928 was detected radio eclipses between the orbital phases $\phi_{\rm in} = 0.18$ and $\phi_{\rm eg} = 0.35$ \citep{yang25}. Assuming isotropic stellar winds, the minimum orbital inclination angle can be estimated to be $i_{\rm min}=90^\circ-360^\circ\times(0.35-0.18)/2=59.4^\circ$, i.e., $\sin i\approx0.86-1.0$. In Figure 1, we plot the upper ($\sin i=1$) and lower limits ($\sin i=0.86$) of the NS mass for different companion-star masses. When $M_{\rm c}=1.4$ and $1.0~M_\odot$, the NS mass of J1928 is in the range of $1.33-2.02~M_\odot$ and $0.65-1.07~M_\odot$, respectively. For a massive companion star of $1.6~M_\odot$, the NS mass can be constrained to $1.74-2.58~M_\odot$, implying the possibility of an extremely massive NS. Based on the mass range of known NSs, the companion mass of J1928 should concentrate in a range of $1.0-1.6~M_\odot$, which is consistent with the inferred mass range using the mass function and optical observations \citep{yang25}.

\section{Orbital evolution of J1928 with a WD companion}

By a detailed binary evolution model, \cite{guo25} showed that an NS binary with a $2.2~M_\odot$ He star and an initial orbital period of $0.1~\rm days$ can evolve into an NS binary with a $1.23~M_\odot$ ONe core, which subsequently begins the WD cooling phase. Adopting the wind-loss-rate prescription for extreme He stars given by \cite{jeff10}, \cite{naga26} found that the young proto-WD can produce a wind-loss rate of $10^{-13}-10^{-12}~M_\odot\,\rm yr^{-1}$, which is sufficient to cause the radio eclipses observed in J1928 in a timescale of $10^4-10^5~\rm yr$. Therefore, the companion of J1928 could be a young CO/ONe WD in the cooling phase.

If the companion of J1928 is a WD, the tidal interaction between the NS and the WD is too weak to produce a detectable effect; thus, GW radiation is a unique mechanism extracting orbital angular momentum \footnote{In principle, we should consider Doppler correction, which is the combined effect of the proper motion and the Galactic acceleration. However, this correction is an order of magnitude of $\dot{P}^{\rm D}\sim10^{-15}~\rm s\,s^{-1}$ \citep[see also][]{laza11}, which is two orders of magnitude smaller the observed $\dot{P}$.}. The period derivative of J1928 due to GW radiation can be expressed as
\begin{equation}
\dot{P}_{\rm gr}=-5.1\times10^{-13}\frac{m_{\rm ns}m_{\rm wd}}{(m_{\rm ns}+m_{\rm wd})^{1/3}}\left(\frac{0.1499~\rm days}{P}\right)^{5/3}~\rm s\,s^{-1},
\end{equation}
where $m_{\rm ns}=M_{\rm ns}/1.0~M_\odot$, $m_{\rm wd}=M_{\rm wd}/1.0~M_\odot$, and $P$ are the NS mass, the WD mass, and the orbital period, respectively. Inserting equation (2) into equation (3), the period derivative of J1928 due to GW radiation is
\begin{equation}
\dot{P}_{\rm gr}=-4.0\times10^{-13}\left(2.066m_{\rm c}^2\sin i-\sqrt{\frac{m_{\rm c}^3}{\sin i}}\right)~\rm s\,s^{-1}.
\end{equation}

Figure 1 also shows the upper and lower limits of the period derivative of J1928 due to GW radiation. If the companion star is an extremely massive WD of $M_{\rm wd}=1.4~M_\odot$, GW radiation produces a period derivative of $-(6.8-9.6)\times10^{-13}~\rm s\,s^{-1}$. To cause radio eclipses, the young, massive WD should drive a sufficiently strong wind ($10^{-13}-10^{-12}~M_\odot\rm yr^{-1}$) in a timescale of $10^4-10^5~\rm yr$ \citep{naga26}. However, such a wind-loss rate could only result in a period derivative at least three orders of magnitude smaller than the observed one (see also Section 4.1). Currently, the observed period derivative of J1928 is $(-4.3\pm4.5)\times10^{-13}~\rm s\,s^{-1}$ \citep{yang25}. Due to a large error bar, it is difficult to determine whether GW radiation is the unique mechanism driving the orbital evolution of J1928. 

\begin{figure}
\centering
\includegraphics[width=\columnwidth]{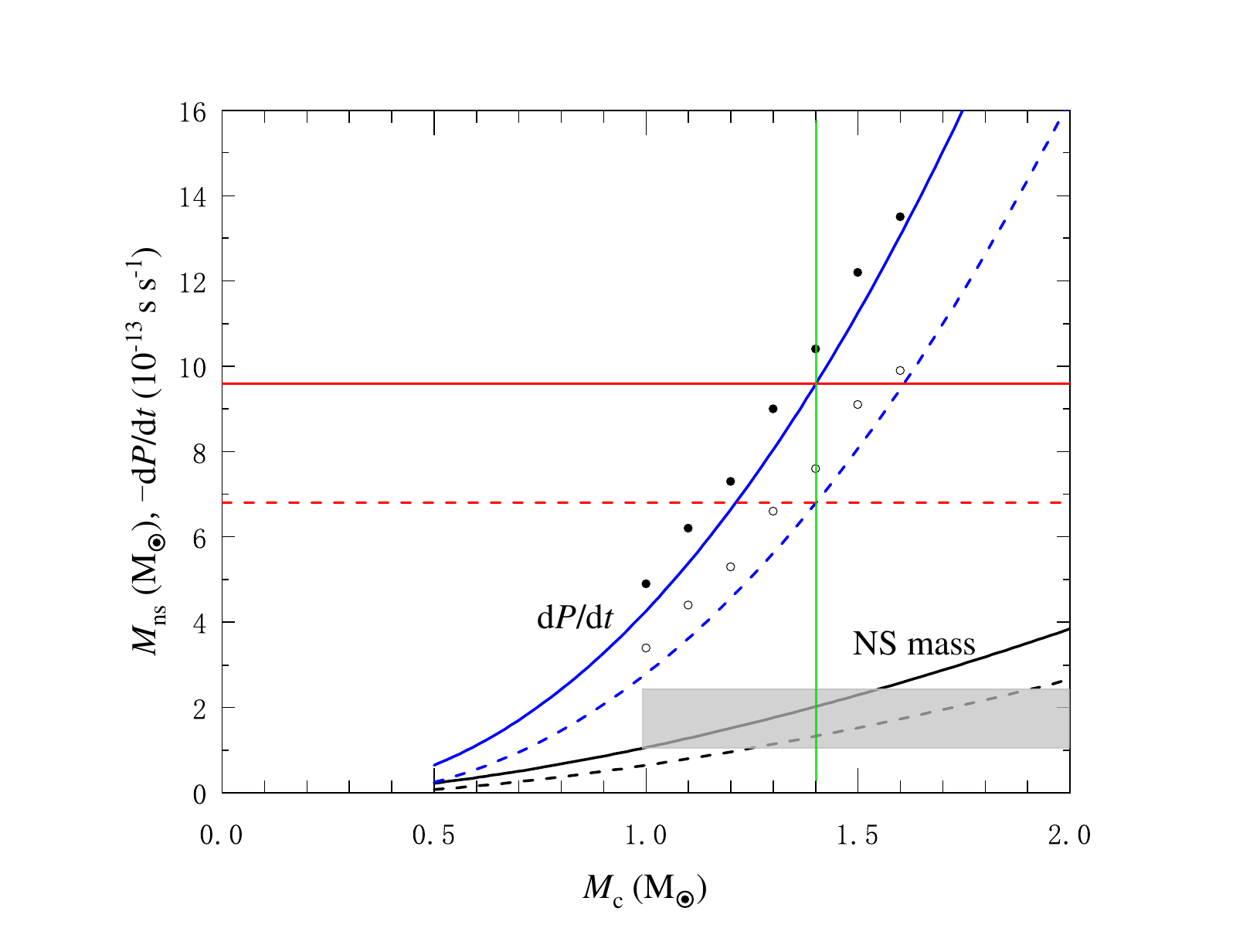}
\caption{NS mass and orbital-period derivative as a function of the companion-star mass for J1928. The black solid and dashed curves describe the NS masses when $\sin i=1$ and $0.86$, respectively. The blue solid and dashed curves represent the orbital-period derivatives when $\sin i=1$ and $0.86$, respectively. The horizontal red solid and dashed lines denote the upper and lower limits of orbital-period derivative when the companion star is a WD, respectively. The green vertical line denotes the upper limit ($1.4~M_\odot$) of WD mass. The solid and open circles correspond to the upper and lower limits of total period derivatives when tidal interactions of He stars are included. The light gray area presents the measured NS mass range of $1.1-2.4~M_\odot$.} \label{fig:orbmass}
\end{figure}

\section{Orbital evolution of J1928 with a He star companion}
\subsection{Orbital evolution of a binary MSP with a He star}

Stripped stars remain a tiny fraction of hydrogen-rich envelope \citep{gotb18}, resulting in relatively large radii and low visual magnitudes. According to the lack of optical and infrared counterpart, a naked He star (with a He abundance of $Y=0.98$ and a metallicity abundance of $Z=0.02$) could be the alternative companion star of J1928. The putative naked He star should be experiencing ferocious nuclear burning because its mass is higher than $0.32~M_\odot$, below which central He burning  would be extinguished \citep{han02}. 

During the evolution of a He star, its envelope expands due to nuclear evolution. According to the angular momentum conservation law, the He star would spin down because of the increasing moment of inertia. However, the tidal interaction between the NS and the He star would continuously accelerate the star up to a synchronous rotation with the orbital motion \citep{patt84}. Therefore, the angular momentum transfers from the orbital motion to the star during its spin-up. As a consequence, the expansion process of a star indirectly extracts the orbital angular momentum \citep{taur99}, and the orbital period derivative will deviate from that fully due to GW radiation if the companion is a He star. Conversely, it is completely contributed by the GW radiation if the companion is a WD.

In a binary MSP consisting of a NS (with a mass $M_{\rm ns}$) and a companion star (with a mass $M_{\rm c}$), the total angular momentum including the orbital angular momentum and the spin angular momentum of the companion is given by
\begin{equation}
J_{\rm tot}=J_{\rm orb}+J_{\rm s}=\mu a^{2}\Omega+I\Omega,
\end{equation}
where $\mu=\frac{M_{\rm ns}M_{\rm c}}{M_{\rm ns}+M_{\rm c}}$ is the induce mass of the system, $a$ is the orbital separation of the system, $\Omega$ is the orbital angular velocity and $I$ is the moment of inertia of the companion star. Considering no mass transfer between the two components and taking the time derivative of the above equation, we can obtain
\begin{equation}
\begin{aligned}
\dot{J}_{\rm tot}=J_{\rm orb}\frac{\dot{M}_{\rm c}}{M_{\rm c}}-J_{\rm orb}\frac{\dot{M}_{\rm c}}{M_{\rm ns}+M_{\rm c}}+J_{\rm orb}\frac{2\dot{a}}{a}\\
+J_{\rm orb}\frac{\dot{\Omega}}{\Omega}
+J_{\rm s}\frac{\dot{I}}{I}+J_{\rm s}\frac{\dot{\Omega}}{\Omega},
\end{aligned}
\end{equation}
where $\dot{M}_{\rm c}$ is the loss rate of stellar winds of the companion star. In general, the winds of the companion star are thought to carry away its specific orbital angular momentum. Therefore, the rate of angular momentum due to stellar winds can be written as
\begin{equation}
\dot{J}_{\rm wind}=J_{\rm orb}\frac{\dot{M}_{\rm c}M_{\rm ns}}{(M_{\rm ns}+M_{\rm c})M_{\rm c}}.
\end{equation}

Because $\dot{J}_{\rm tot}=\dot{J}_{\rm gw}+\dot{J}_{\rm wind}$
(where $\dot{J}_{\rm gw}$ is the rate of angular momentum loss due to GW radiation), equation (6) can be rewritten as
\begin{equation}
\dot{J}_{\rm gw}=J_{\rm orb}\frac{2\dot{a}}{a}
+J_{\rm orb}\frac{\dot{\Omega}}{\Omega}
+J_{\rm s}\frac{\dot{I}}{I}+J_{\rm s}\frac{\dot{\Omega}}{\Omega}.
\end{equation}
From Kepler's third law ($G(M_{\rm ns} + M_{\rm c})/a^3 = \Omega^{2}$), we have
\begin{equation}
\frac{\dot{M}_{\rm c}}{M_{\rm ns} + M_{\rm c}}-3\frac{\dot{a}}{a}=2\frac{\dot{\Omega}}{\Omega}.
\end{equation} 
Inserting equation (9) into equation (8), it yields
\begin{equation}
\dot{J}_{\rm gw}=J_{\rm orb}\frac{2\dot{M}_{\rm c}}{3(M_{\rm ns} + M_{\rm c})}
-J_{\rm orb}\frac{\dot{\Omega}}{3\Omega}
+J_{\rm s}\frac{\dot{I}}{I}+J_{\rm s}\frac{\dot{\Omega}}{\Omega}.
\end{equation}

To produce the radio eclipses observed in 17\% orbital phase of J1928, the wind momentum ratio between the companion star and the MSP is $\eta_{\rm w}=0.016$ \citep{naga26}. According to the spin-down luminosity of J1928 is $L_{\rm sd}=1.2\times10^{35}~\rm erg\,s^{-1}$ \citep{yang25}, the wind loss rate of the He star can be estimated to be $\dot{M}_{\rm c}=\eta_{\rm w}L_{\rm sd}/(cv_{\rm w})\approx5.0\times10^{-12}~M_\odot\rm yr^{-1}$, where $v_{\rm w}=2.0\times10^8~\rm cm\,s^{-1}$ is the wind speed (i.e. the escape velocity of the He star). Thus, $2\dot{M}_{\rm c}/(M_{\rm ns}+M_{\rm c})\sim-5.0\times10^{-12}~\rm yr^{-1}$ in equation (10).
However, the GW radiation can produce a minimum period-change rate as $\dot{P}/P=-\dot{\Omega}/\Omega=-6.8\times10^{-10}~\rm yr^{-1}$ if the companion of J1928 is a He star with $M_{\rm c}=1.0~M_\odot$, which is $2$ orders of magnitude higher than that contributed by the wind loss of the He star. Therefore, we ignore the contribution of stellar winds on the orbital evolution of J1928.

\subsection{Constraining the evaporation efficiency of spin-down luminosity}

Actually, the irradiation process on the surface of the He star of the spin-down luminosity ($L_{\rm sd}$) of the energetic MSP can also produce evaporation winds. The loss rate of evaporation winds is given by
\begin{equation}
\dot{M}_{\rm ev,w}= \frac{f_{\rm ev}L_{\rm sd}R_{\rm c}^{3}}{4GM_{\rm c}a^{2}},
\end{equation}
where $G$ is the gravitational constant, $f_{\rm ev}$ is the evaporation efficiency, and $R_{\rm c}$ is the radius of the companion star. From Kepler's third law, the orbital separation of PSR J1928+1815 is
\begin{equation}
a=\left(\frac{G(M_{\rm ns}+M_{\rm c})P^{2}}{4\pi^{2}}\right)^{1/3}=1.1\times10^{11}~{\rm cm}\sqrt{m_{\rm c}\sin i},
\end{equation}
where $m_{\rm c}=M_{\rm c}/M_\odot$. Furthermore, a stellar evolution model using MESA (version 12115) obtained a radius-mass relation of naked He stars as  \citep{yang25}
\begin{equation}
R_{\rm He}=0.205R_\odot~m_{\rm He}^{0.786} 
\end{equation}
Inserting equations (12) and (13) into equation (11), the evaporation efficiency of spin-down luminosity can be written as
\begin{equation}
\begin{aligned}
f_{\rm ev}\approx0.006\times\left(\frac{\dot{M}_{\rm ev,w}}{5.0\times10^{-12}~M_\odot\,\rm yr^{-1}}\right)
\left(\frac{1.2\times10^{35}~\rm erg\,s^{-1}}{L_{\rm sd}}\right)\\
\times\frac{\sin i}{m_{\rm c}^{0.358}}.
\end{aligned}
\end{equation}

According to the observed radio eclipses, the wind-loss rate is estimated to be $\dot{M}_{\rm c}\approx5.0\times10^{-12}~M_\odot\rm yr^{-1}$ in the He star case. Considering the intrinsic winds of the He star, the loss rate of evaporation winds should be $\dot{M}_{\rm ev,w}<5.0\times10^{-12}~M_\odot\rm yr^{-1}$. Therefore, the upper limit of the evaporation efficiency in J1928 is $f_{\rm ev,max}\approx 0.006$ when $M_{\rm c}=1.0~M_\odot$ and $\sin i=1$. This evaporation efficiency is an order of magnitude smaller than those ($f_{\rm ev}=(0.26-4.5)\times10^{-4}$) in black widows derived by \cite{ginz20}. However, \cite{ginz20} only considered the evaporation process of $\gamma$-luminosity of MSPs. In general, about 10\% of a MSP's spin-down luminosity converts into $\gamma$-rays energy in the band of 0.1-100 GeV \citep{abdo13}. Therefore, our estimated evaporation efficiency is approximately consistent with those derived by \cite{ginz20}.

\subsection{Orbital evolution when stellar winds of He stars are ignored}
Therefore, we ignore the first and second terms on the right-hand side of equation (6), Due to $\dot{P}/P=-\dot{\Omega}/\Omega$, and equation (10) can be modified as
\begin{equation}
\dot{J}_{\rm gw}=J_{\rm orb}\frac{\dot{P}}{3P}
+J_{\rm s}\frac{\dot{I}}{I}+J_{\rm s}\frac{\dot{\Omega}}{\Omega}.
\end{equation}
When the tidal effect is not included, equation (15) simplifies to
\begin{equation}
\dot{J}_{\rm gw}=J_{\rm orb}\frac{\dot{P}}{3P}.
\end{equation}
Therefore, the period derivative due to GW radiation is
\begin{equation}
\dot{P}_{\rm gw}=\frac{3\dot{J}_{\rm gw}}{J_{\rm orb}}P.
\end{equation}

When the tides effect is considered, from equation (15) the total period derivative is derived as
\begin{equation}
\dot{P}_{\rm total}=\frac{\dot{P}_{\rm gw}-3rP\dot{I}/I}{1-3r},
\end{equation}
where $r=J_{\rm s}/J_{\rm orb}$ is the ratio between the spin angular momentum of the He star and the orbital angular momentum. Taking $M_{\rm c}=1.0~M_\odot$, the minimum mass of the NS is $M_{\rm ns,min}=0.65~M_\odot$ for J1928. The minimum orbital angular momentum is $J_{\rm orb,max}\approx3.7\times10^{51}~\rm g\,cm^2s^{-1}$, but the spin angular momentum of the He star is $J_{\rm s}\approx1.5\times10^{49}~\rm g\,cm^2s^{-1}$. Since the maximum ratio between $\dot{J}_{\rm s}$ and $\dot{J}_{\rm orb}$ is $r_{\rm max}=4.1\times10^{-3}$, we ignore the term of $3r$ in the denominator of equation (18). Therefore, the joint contribution of the tidal interaction and GW radiation produce a total period derivative as
\begin{equation}
\begin{aligned}
\dot{P}_{\rm total}\approx\dot{P}_{\rm gw}+\dot{P}_{\rm tide}=\dot{P}_{\rm gw}-3rP\dot{I}/I
=\dot{P}_{\rm gw}-\\ 7.4\times10^{-14}~{\rm s\,s^{-1}}\times
\left(\frac{r}{3.0\times10^{-3}}\right)\left(\frac{\dot{I}/I}{2.0\times10^{-8}~\rm yr^{-1}}\right).
\end{aligned}
\end{equation}

According to equation (4), the maximum orbital period derivative due to GW radiation for J1928 with a He star companion of $1.0-1.6~M_\odot$  is $\dot{P}_{\rm gw}=-(4.3-13.1)\times10^{-13}~\rm s\,s^{-1}$. However, the orbital period derivative of the same system is $-(5.0-13.8)\times10^{-13}~\rm s\,s^{-1}$ when the tidal interaction is included, and $\dot{I}/I=2.0\times10^{-8}~\rm yr^{-1}$ and $r=3.0\times10^{-3}$. Because the maximum period derivative of a binary MSP consisting of a NS and a WD is $-9.6\times10^{-13}~\rm s\,s^{-1}$, the possibility that the companion star of J1928 is a WD can be ruled out if its measured period derivative is in the range of $-(9.6-13.6)\times10^{-13}~\rm s\,s^{-1}$.

\subsection{Stellar evolution models of He stars}
We utilize a binary update version of the Modules for Experiments in Stellar Astrophysics (MESA) code \citep[version r-12115, ][]{paxt11,paxt13,paxt15,paxt18,paxt19} to model the evolution of He stars' inertia of moment. We build a zero age MS naked He star model with solar metallicity \citep{wong19}, which consists of 98\% He and 2\% metallicity (i.e., $Y = 0.28, Z = 0.02$).  Once the donor star starts He burning, Type 2 opacities are used for extra C/O burning. The MESA code continuously models the evolution of He stars with different initial masses until He stars begin to contract, i.e., $\dot{I}<0$.

Figure 2 shows the evolution of $\dot{I}/I$ of He stars with different initial masses. When He stars' ages are $1-3~\rm Myr$, the relative change rates of inertia of moments are in the range of $\dot{I}/I\sim(1-4)\times10^{-8}~\rm yr^{-1}$.

Table 1 summarizes the typical simulated parameters of naked He stars at the age of 2 Myr. The minimum ratios between the spin angular momentum of the He star and the orbital angular momentum are $r_{\rm min}\approx0.003$, and the relative change rates of inertia of moments are $\dot{I}/I\sim(0.9-2.5)\times10^{-8}~\rm yr^{-1}$. The period derivatives due to GR are in the range of $-(4.3-13.1)\times10^{-13}~\rm s\,s^{-1}$, but the total period derivative are $\dot{P}_{\rm total}=-(4.9-13.5)\times10^{-13}~\rm s\,s^{-1}$ if the one ($\dot{P}_{\rm tide}=-(0.4-1.0)\times10^{-13}~\rm s\,s^{-1}$) induced by the tidal effect is included. Therefore, the tidal interaction between an NS and a He star can enhance the period derivative by a factor of $\sim 10\%$. In Figure 1, we also show the maximum and minimum $\dot{P}_{\rm total}$ when the companion of J1928 is a He star. It is clear that the maximum $\dot{P}_{\rm total}$ in $1.4$, $1.5$, and $1.6~M_\odot$ He stars cases and the minimum $\dot{P}_{\rm total}$ in $1.6~M_\odot$ He star case exceed the upper limit of $\dot{P}_{\rm gw}$ in $1.4~M_\odot$ WD companion case.

Figure 3 plots the two period derivatives ($\dot{P}_{\rm total}$ and $\dot{P}_{\rm gw}$) under different orbital inclination angles. Because of the uncertainty of orbital inclination angle, the period derivative of a binary MSP with a He star of $1.4~M_\odot$ is unlikely to exceed the upper limit ($-9.6\times10^{-13}~\rm s\,s^{-1}$) of $\dot{P}_{\rm gw}$ of J1928 with a WD companion. If the tidal effect is included, the predicted period derivatives are possible to exceed this upper limit when the He-star mass is in the range of $1.35-1.5~M_\odot$ (the orbital inclination angle has to satisfy $\sin i\sim0.89$, $0.96$, and 0.99 when $M_{\rm he}=1.5$, $1.4$, and $1.35~M_\odot$, respectively). Therefore, it is possible to confirm a He star with a mass of $1.35-1.5~M_\odot$ in a high orbital inclination angle if the tidal effect is included.

\begin{table*}
\begin{center}
\caption{Some typical simulated  parameters of naked He stars \label{tbl-2}}
\begin{tabular}{@{}lllllllll@{}}
\hline\hline\noalign{\smallskip}
He-star mass& radius  & $I$ & $(\mu a^2)_{\rm max}$ &$r_{\rm min}$ & $\dot{I}/I$ &$\dot{P}_{\rm gw,max}$  &$\dot{P}_{\rm tide,min}$ & $\dot{P}_{\rm total}$\\
($M_\odot$)& ($R_\odot$)& ($10^{52}~\rm g\,cm^2$)& ($10^{55}~\rm g\,cm^2$)& & ($10^{-8}~\rm yr^{-1}$)& ($10^{-13}~\rm s\,s^{-1}$) & ($10^{-13}~\rm s\,s^{-1}$)&($10^{-13}~\rm s\,s^{-1}$)\\
\hline\noalign{\smallskip}
1.0  & 0.215  &3.0  & 1.1 & 0.003  & 1.6  & $-4.3$  &  $-0.6$ & $-4.9$  \\
1.1 & 0.233  & 3.9  & 1.4 & 0.003  & 2.1  & $-5.4$  &  $-0.8$ & $-6.2$ \\
1.2 & 0.251  & 5.0  & 1.8 & 0.003  & 1.8  & $-6.6$  &  $-0.7$ & $-7.3$ \\
1.3 & 0.269  & 6.2  & 2.1 & 0.003  & 2.5  & $-8.0$  &  $-1.0$ & $-9.0$ \\
1.4 & 0.287  &7.5   & 2.5 & 0.003  & 1.9  & $-9.6$  &  $-0.8$ & $-10.4$ \\
1.5 & 0.304  &9.0   & 3.0 & 0.003  & 2.5  & $-11.2$  &  $-1.0$ &$-12.2$ \\
1.6 & 0.321  &10.6  & 3.5 & 0.003  &0.9   & $-13.1$  &  $-0.4$ &$-13.5$ \\
\hline\noalign{\smallskip}
\end{tabular}
\\
\end{center}
\label{tl}
\end{table*}

\subsection{Searching for the He star in near and mid-infrared bands}

A detailed stellar evolution model using MESA derived a radius-mass relation as $R_{\rm he}=0.205~R_\odot (M_{\rm he}/M_\odot)^{0.786}$ and an effective temperature-mass relation as $T_{\rm eff}=49600(M_{\rm he}/M_\odot)^{0.495}~\rm K$ for naked He stars with $Y=0.98$ and $Z=0.02$ at an age of $1 ~\rm Myr$ \citep{yang25}. We first calculate the intrinsic fluxes of He stars in different bands according to Planck's equation, then derive the corresponding detected fluxes $f_{\nu}$. Subsequently, we obtain the magnitudes of He stars with masses of $1.4-1.6~M_\odot$ adopting the Vega magnitude system \citep{bess98}. Based on the dispersion measure detected in J1928, its distance was inferred to 9.7 or 7.2 kpc using two Galactic electron density distribution models \citep{yang25}. In table 2, we summarize the estimated Vega magnitudes and fluxes of naked He stars in three bands including $K_s$, W1 and W2 of Wide-field Infrared Survey Explorer (WISE). WISE achieved at least 80 and $110~\mu \rm Jy$ $5\sigma$ sensitivity in W1 ($3.4~\rm \mu m$) and W2 ($4.6~\rm \mu m$) bands, which correspond to Vega magnitudes of 16.5 and 15.5 \citep{wrig10}, respectively. According to the magnitude and flux limits, the mid-infrared survey of WISE cannot detect a naked He star with a mass of $1.4-1.6~M_\odot$ at a distance of 9.7 or 7.2 kpc. However, the deep near-infrared imaging using Keck/NIRC2 rules out some candidates of naked He stars \citep{naga26}. Even if there exists an extinction with four magnitudes in the $K_s$
band \citep{lewi23}, a He star with a mass of $1.5-1.6~M_\odot$ as the companion of J1928 can be fully ruled out if the distance is $7.2~\rm kpc$. When $d=9.7~\rm kpc$, a He star with a mass of $1.4-1.6~M_\odot$ is still likely to be the companion of J1928 if the extinction exceeds 3.5 magnitudes.

\begin{table}
\begin{center}
\caption{Estimated Vega magnitudes and fluxes of naked He stars in different bands\label{tbl-2}}
\begin{tabular}{@{}lllll@{}}
\hline\hline\noalign{\smallskip}
Parameters &  & naked & & observation\\
           &  & He stars& & limits\\
\hline\noalign{\smallskip}
$M_{\rm he}$ ($M_\odot$) & 1.4 & 1.5 &  1.6  &   $\cdots$    \\
$R_{\rm he}$ ($R_\odot$) & 0.267  &  0.282   &  0.297    & $\cdots$        \\
$T_{\rm eff}$  ($\rm K$) & 58600  &  60600   & 62600     &  $\cdots$        \\
\hline\noalign{\smallskip}
&   & $d=9.7~\rm kpc$  &    &     \\
$\rm K_{s}$ (mag) & 18.1 & 17.9  &17.8 &  21.3 \\
$\rm W1$ (mag) & 18.2 & 18.0 & 17.9 & 16.5 \\
$\rm W2$ (mag) & 18.2 & 18.0 & 17.9  &  15.5 \\
$f_{\rm \nu, W1}$ ($\rm \mu Jy$)& 16.4  & 19.0  & 21.8 & 80 \\
$f_{\rm \nu, W2}$ ($\rm \mu Jy$)& 9.0  & 10.4   & 12.0 &  110 \\
\hline\noalign{\smallskip}
&   & $d=7.2~\rm kpc$  &    &     \\
$\rm K_{s}$ (mag) & 17.4 & 17.3  &17.1 &  21.3 \\
$\rm W1$ (mag) & 17.5 & 17.4 & 17.2 & 16.5 \\
$\rm W2$ (mag) & 17.5 & 17.4 & 17.2  &  15.5 \\
$f_{\rm \nu, W1}$ ($\rm \mu Jy$)& 29.7  & 34.4  & 39.3 & 80 \\
$f_{\rm \nu, W2}$ ($\rm \mu Jy$)&16.4 & 19.0   & 21.7 &  110 \\
\hline\noalign{\smallskip}
\end{tabular}
\\
\end{center}
\label{tl}
\end{table}

\begin{figure}
\centering
\includegraphics[width=\columnwidth]{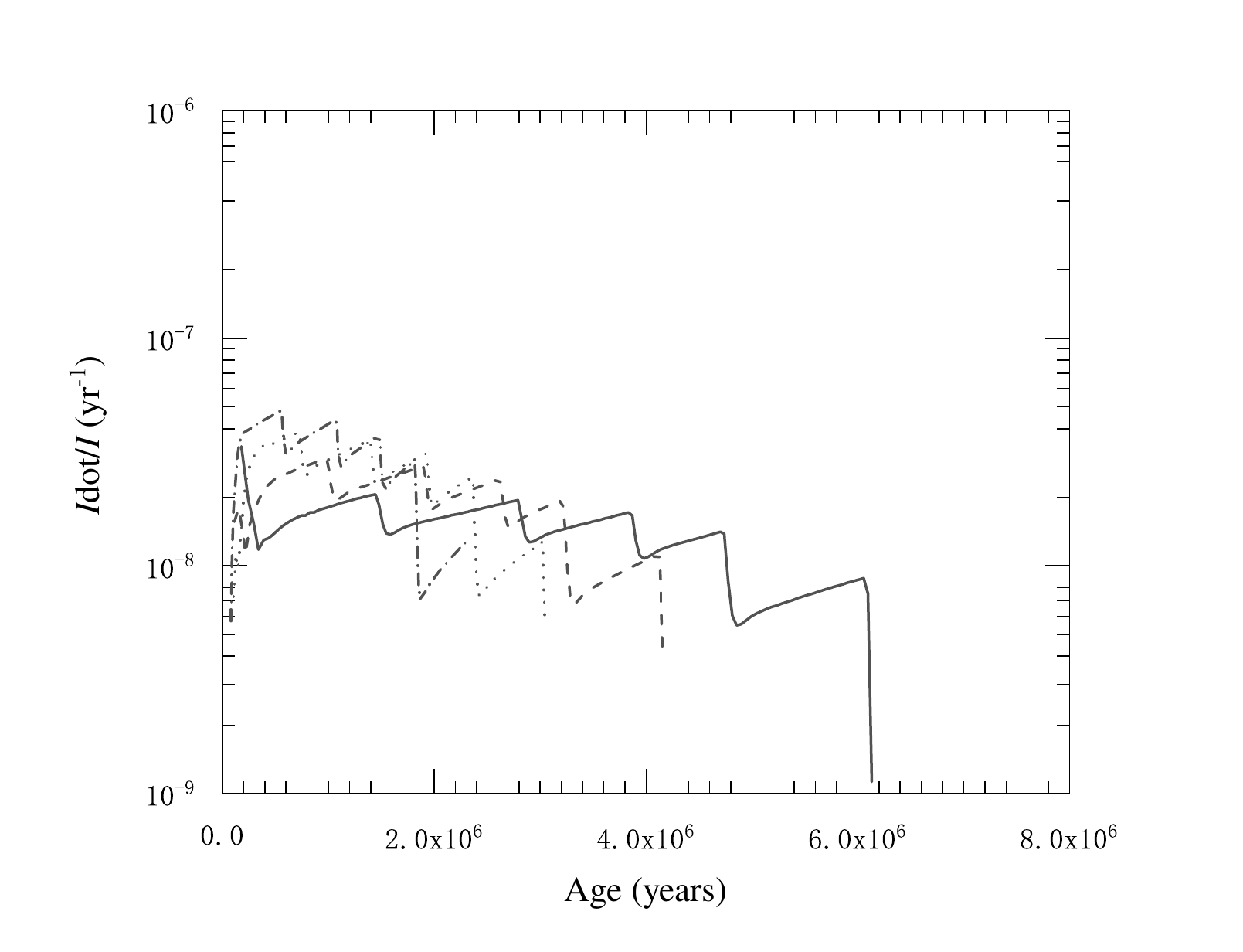}
\caption{Evolution of $\dot{I}/I$ of naked He stars as a function of their ages. The solid, dashed, dotted, and dashed-dotted curves represent the evolutionary tracks of He stars with $M_{\rm he}=1.0, 1.2, 1.4$, and $1.6~M_{\odot}$, respectively.} 
\end{figure}

\begin{figure}
\centering
\includegraphics[width=\columnwidth]{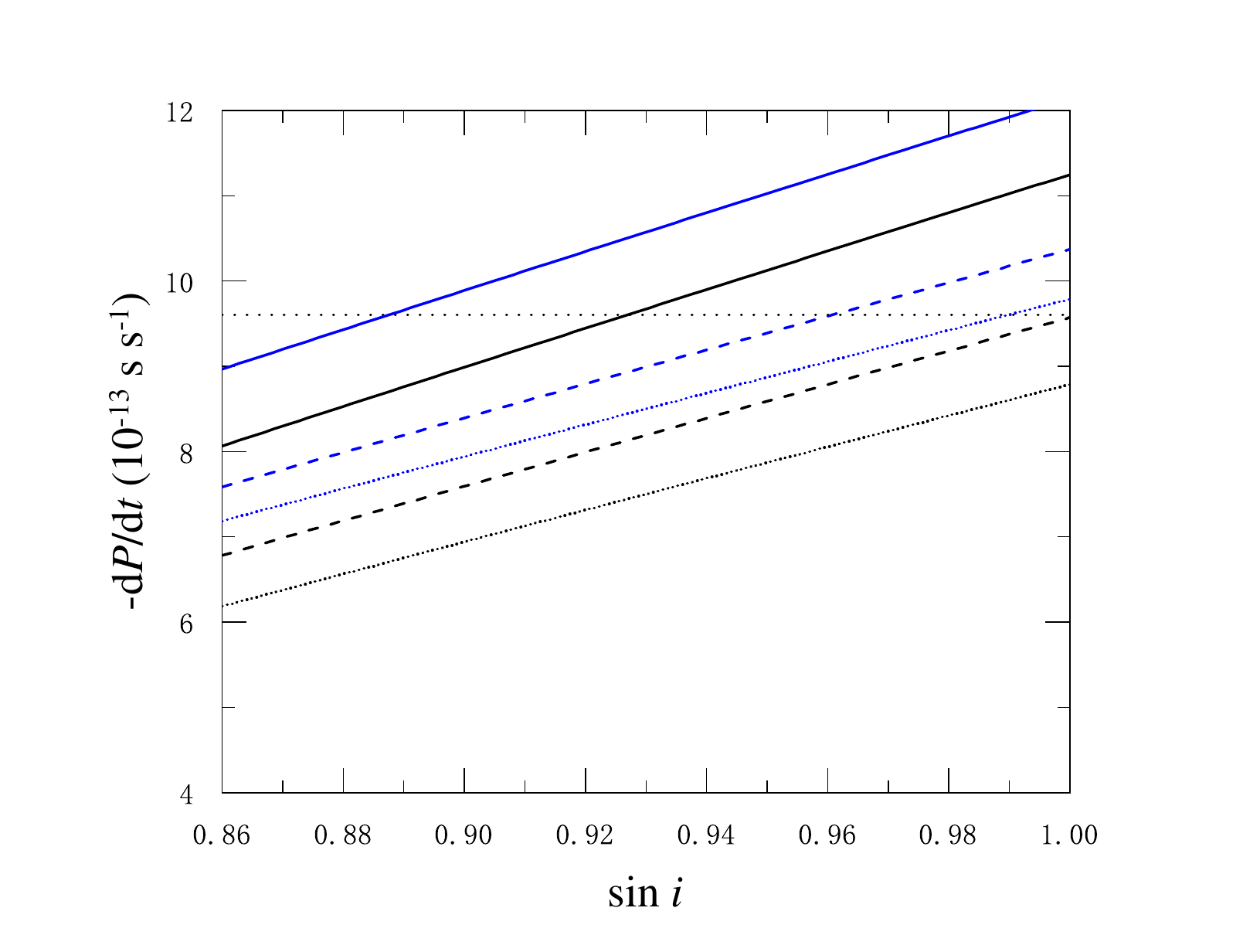}
\caption{Orbital-period derivative as a function of $\sin i$ when the companion star of J1928 is a He star. The solid, dashed, and dotted curves correspond to the two cases with $M_{\rm he}=1.5$, $1.4$, and $1.35~M_\odot$, respectively. The blue and black curves represent $\dot{P}_{\rm total}$ and $\dot{P}_{\rm gw}$, respectively. The horizontal dotted line denotes the upper limit ($-9.6\times10^{-13}~\rm s\,s^{-1}$) of $\dot{P}_{\rm gw}$ when the companion star of J1928 is an extremely massive WD.} 
\end{figure}

\section{Discussion and Summary}
At present, the companion type of J1928 still remains unclear. The population synthesis simulation predicts that numbers of J1928-like systems with He stars are three orders of magnitude higher than those for the pre-WD case in the Galaxy \citep{yang25}. However, no detection through deep near-infrared imaging using Keck/NIRC2 suggested that the companion star of J1928 could be a pre-WD \citep{naga26}. Therefore, J1928 is undoubtedly a rare spider pulsar, regardless of whether the companion is a He star or a WD. 

Due to uncertainties in extinction and distance, no detection of NIR imaging observation with limiting magnitudes of $K_{\rm s}=21.3$ and the mid-infrared survey of WISE cannot fully rule out the possibility that the companion of J1928 is a He star. We propose an indirect way to test the companion nature of J1928 in the observation. Based on the mass function and orbital inclination angle inferred by the radio eclipses, the gravitational radiation can produce a period derivative of $-(6.8-9.6)\times10^{-13}~\rm s\,s^{-1}$ for J1928 with an extremely massive ($1.4~M_\odot$) WD companion. Therefore, the companion star of J1928 should be a He star with a mass higher than $1.4~M_\odot$ if its measured period derivative exceeds $-9.6\times10^{-13}~\rm s\,s^{-1}$. The tidal interaction between an NS and a He star is much stronger than that between an NS and a WD. The expansion of the He star during its evolution could transfer angular momentum from the orbital motion to the spin motion due to synchronous rotation. Therefore, tidal interactions could produce an additional contribution to the period derivative due to gravitational radiation. Our detailed stellar evolution models indicate that the relative change rates of He stars' moment of inertia are $\dot{I}/I\approx(0.9-2.5)\times10^{-8}~\rm yr^{-1}$, which result in an additional period derivative of $\dot{P}_{\rm tide}\approx-(0.4-1.0)\times10^{-13}~\rm s\,s^{-1}$ and a total one of $-(4.9-13.5)\times10^{-13}~\rm s\,s^{-1}$ when the He-star mass is in the range of $1.0-1.6~M_\odot$. Especially, the tidal interaction could help to confirm a He star with a mass of $1.35-1.5~M_\odot$, depending on the orbital inclination angle.  Meanwhile, the maximum evaporation efficiency of spin-down luminosity is constrained to be $0.006$ when the companion star of J1928 is a He star.

So far, the most massive pulsating WD J0049-2525 was inferred to have a mass of $\sim1.3~M_\odot$ \citep{cali25}. GW radiation can produce a maximum period derivative of $-8.0\times10^{-13}~\rm s\,s^{-1}$ for J1928 with a companion star of $1.3~M_\odot$. Therefore, J1928 should possess an extremely massive WD companion or a He star companion if the measured period derivative exceeds $-8.0\times10^{-13}~\rm s\,s^{-1}$.


In theory, a third body can also produce an additional orbital-period derivative. For example, a $M_3=1.0~M_\odot$ sun-like star in a outer orbit of $3000~\rm years$ (taking the distance between the third body and the inner binary $d=300~\rm AU$) can cause a maximum orbital-period derivative as $\dot{P}=GM_3P/(d^2c)\approx2.8\times10^{-12}~\rm s\,s^{-1}$. Recently, the binary MSP PSR J0435+3233 was identified to have a Sun-like star as the distant tertiary \citep{yang26}. However, the existence of a third body can cause synchronous changes of spin-period and orbital-period derivatives. A long-term timing observation for J1928 could confirm or negate the existence of a third body.

\begin{acknowledgements}
We appreciate the referee for a very careful reading and constructive comments that have led to the improvement of the manuscript. This work was partly supported by the National Natural Science Foundation of China (under grant Nos. 12673038 and 12273014), and the Natural Science Foundation (under grant number ZR2025MS43) of Shandong Province.   
\end{acknowledgements}

\end{document}